\documentclass[12pt]{article}
\usepackage{amssymb}
\usepackage{epsfig}
\usepackage{amsmath}
\newtheorem{theorem}{Theorem}

\begin{document}

\title{An infinite-dimensional calculus for gauge theories}
\author{R. Vilela Mendes \thanks{%
IPFN - EURATOM/IST Association, Instituto Superior T\'{e}cnico, Av. Rovisco
Pais 1, 1049-001 Lisboa, Portugal, http://label2.ist.utl.pt/vilela/} \thanks{%
CMAF, Complexo Interdisciplinar, Universidade de Lisboa, Av. Gama Pinto, 2 -
1649-003 Lisboa (Portugal), e-mail: vilela@cii.fc.ul.pt}}
\date{}
\maketitle
\tableofcontents

\begin{abstract}
A space for gauge theories is defined, using projective limits as subsets of
Cartesian products of homomorphisms from a lattice on the structure group.
In this space, non-interacting and interacting measures are defined as well
as functions and operators. From projective limits of test functions and
distributions on products of compact groups, a projective gauge triplet is
obtained, which provides a framework for the infinite-dimensional calculus
in gauge theories. The gauge measure behavior on nongeneric strata is also
obtained.
\end{abstract}

\section{Preliminaries}

\subsection{Gauge theory. Basic definitions}

In its usual formulation a classical gauge theory consists of four basic
objects:

(i) A principal fiber bundle $P\left( M,G\right) $ with structure group $G$
and projection $\pi :P\rightarrow M$, the base space $M$ being an oriented
Riemannian manifold.

(ii) An affine space $\mathcal{C}$ of connections $\omega $ on $P$, modelled
by a vector space $\mathcal{A}$ of 1-forms on $M$ with values on the Lie
algebra $LG$ of $G$.

(iii) The space of differentiable sections of $P$, called the \textit{gauge
group }$\mathcal{G}$

(iv) A $\mathcal{G}-$invariant functional (the Lagrangian) $\mathcal{L}:%
\mathcal{A}\rightarrow \mathbb{R}$

The statement (iv) presumes the existence of a reference measure in the
configuration space $\mathcal{A}/\mathcal{G}$, the exponential of the
Lagrangian being a Radon-Nykodim derivative with respect to this reference
measure. Because that might not always be possible to achieve, it is better
to replace (iv) by:

(iv$^{\prime }$) A well defined measure in the configuration space $\mathcal{%
A}/\mathcal{G}$.\ 

Choosing a reference connection, the affine space of connections on $P$ may
be modelled by a vector space of $LG$-valued 1-forms ($C^{\infty }\left(
\Lambda ^{1}\otimes LG\right) $). Likewise the curvature $F$ is identified
with an element of $C^{\infty }\left( \Lambda ^{2}\otimes LG\right) $. In a
coordinate system one writes 
\begin{equation*}
A=A_{\mu }^{a}dx^{\mu }t_{a}\qquad x\in M\qquad t_{a}\in LG
\end{equation*}%
with the action of $\gamma =\{g\left( x\right) \}\in \mathcal{G}$ on $A\in 
\mathcal{A}$ given by 
\begin{equation}
\gamma :A_{\mu }\left( x\right) \rightarrow \left( gA_{\mu }\right) \left(
x\right) =g\left( x\right) A_{\mu }\left( x\right) g^{-1}\left( x\right)
-\left( \partial g\right) \left( x\right) \cdot g^{-1}\left( x\right)
\label{2.4}
\end{equation}%
In this paper $G$ will always be considered to be a compact group.

The action of $\mathcal{G}$ on $\mathcal{A}$ leads to a stratification of $%
\mathcal{A}$ corresponding to the classes of equivalent \textit{orbits} $%
\left\{ gA;g\in \mathcal{G}\right\} $. Let $S_{A}$ denote the \textit{%
isotropy (or stabilizer) group} of $A\in \mathcal{A}$%
\begin{equation}
S_{A}=\left\{ \gamma \in \mathcal{G}:\gamma A=A\right\}  \label{2.5}
\end{equation}
The \textit{stratum} $\Sigma \left( A\right) $ of $A$ is the set of
connections having isotropy groups $\mathcal{G}-$conjugated to that of $A$%
\begin{equation}
\Sigma \left( A\right) =\left\{ B\in \mathcal{A}:\exists \gamma \in \mathcal{%
G}:S_{B}=\gamma S_{A}\gamma ^{-1}\right\}  \label{2.6}
\end{equation}
The \textit{configuration space of the gauge theory} is the quotient space $%
\mathcal{A}/\mathcal{G}$ and therefore a stratum is the set of points in $%
\mathcal{A}/\mathcal{G}$ that correspond to orbits with conjugated isotropy
groups.

\subsection{Generalized connections and the Ashtekar-Lewandowski measure}

Whenever a Lagrangian is defined the calculation of physical quantities in
the path integral formulation 
\begin{equation}
\left\langle \phi \right\rangle =\int_{\mathcal{A}/\mathcal{G}}\phi \left(
\xi \right) e^{-\mathcal{L}\left( \xi \right) }d\mu \left( \xi \right)
\label{2.7}
\end{equation}%
requires a measure in $\mathcal{A}/\mathcal{G}$ and no such measure is found
for Sobolev connections. Therefore it turned out to be more convenient to
work in a space of \textit{generalized connections} $\overline{\mathcal{A}}$%
, defining parallel transports on piecewise smooth paths as simple
homomorphisms from the paths on $M$ to the group $G$, without a smoothness
assumption\cite{Ashtekar1}. The same applies to the generalized gauge group $%
\overline{\mathcal{G}}$. Then, there is in $\overline{\mathcal{A}}/\overline{%
\mathcal{G}}$ an induced Haar measure, the Ashtekar-Lewandowski (AL) measure%
\cite{Ashtekar2} \cite{Ashtekar3}. Sobolev connections are a dense zero
measure subset of the generalized connections\cite{Marolf}.

A generalized connection is simply a homomorphism from the groupoid of paths 
$\mathcal{P}$ to the structure group. Different choices for the groupoid of
paths have been proposed. Piecewise analyticity was first proposed \cite%
{Ashtekar0}, later extended to the smooth category\cite{Baez1} and to a more
general setting covering both cases \cite{Fleichhack1} \cite{Fleischhack2}.
This led to the notions of graphs\cite{Ashtekar2}, webs and hyphs\cite%
{Fleichhack2}. Here, being mostly concerned with the Yang-Mills theory (and
not with gravity where diffeomorphism invariance is important), the
piecewise analytic case will be considered.

The space of generalized connections is characterized by using the
representation theory of $C^{\ast }$ algebras. The relevant algebra is the
algebra $\mathcal{HA}$ of functions on $\mathcal{A}/\mathcal{G}$ obtained by
taking finite linear combinations of finite products of traces of holonomies
(Wilson loop functions $W_{\alpha }\left( A\right) $) around closed loops $%
\alpha $. The completion $\overline{\mathcal{HA}}$ of $\mathcal{HA}$ in the
sup norm is a $C^{\ast }$ algebra. $\overline{\mathcal{HA}}$ is an abelian $%
C^{\ast }$ algebra and its Gel'fand spectrum is, by definition, the space of
generalized connections $\overline{\mathcal{A}/\mathcal{G}}$ . An
equivalence class of holonomically equivalent loops is called a \textit{hoop}%
. By composition, hoops generate a hoop group $\mathcal{HG}$. Every point $%
\overline{A}\in \overline{\mathcal{A}/\mathcal{G}}$ gives rise to a
homomorphism $\overset{\backsim }{H}$ from the hoop group $\mathcal{HG}$ to
the structure group $G$ and every homorphism defines a point in $\overline{%
\mathcal{A}/\mathcal{G}}$ and, from $\overset{\backsim }{H}_{\overline{A}%
}\left( \beta \right) =TrH\left( \beta ,\overline{A}\right) $, where $%
H\left( \beta ,\overline{A}\right) $ denotes the holonomy, it follows that
the correspondence has the trivial ambiguity that $\overset{\backsim }{H}$
and $g^{-1}\overset{\backsim }{H}g$ define the same point in $\overline{%
\mathcal{A}/\mathcal{G}}$ \cite{Ashtekar0}.

An important notion is the notion of independent hoops. Denote by $\overset{%
\backsim }{\alpha }$ the hoop for which $\alpha $ is a representative loop.
In a set of independent hoops $\left\{ \overset{\backsim }{\beta }_{1},%
\overset{\backsim }{\beta }_{2},\cdots ,\overset{\backsim }{\beta }%
_{n}\right\} $ every representative loop $\beta _{i}$ must contain an open
interval that is traversed exactly once and no finite segment of which is
shared by any other loop in a different hoop. Furthermore given a set of
hoops $\left\{ \overset{\backsim }{\alpha }_{1},\overset{\backsim }{\alpha }%
_{2},\cdots ,\overset{\backsim }{\alpha }_{k}\right\} $ it is always
possible to find a set of independent hoops such that the hoop subgroup
generated by the $\overset{\backsim }{\alpha }_{i}$'s is contained in the
hoop subgroup generated by the $\overset{\backsim }{\beta }_{i}$'s and for
every $\left( g_{1},g_{2},\cdots ,g_{n}\right) \in G^{n}$ there is a
connection $A\in \mathcal{A}$ such that $H\left( \beta _{i},A\right) =g_{i}$
, $i=1,\cdots ,n$ \cite{Ashtekar0}. $H\left( \beta _{i},A\right) $ denotes
the holonomy $\left( W_{\beta }\left( A\right) =TrH\left( \beta
_{i},A\right) \right) $.

The notion of independent hoops provides a simple definition of cylindrical
functions. Given a set of independent loops $\beta _{1},\beta _{2},\cdots
,\beta _{n}$, consider the hoop subgroup $S^{\ast }$ that they generate and
define an equivalence relation in $\overline{\mathcal{A}/\mathcal{G}}$ by $%
\overline{A}_{1}\sim \overline{A}_{2}$ iff $\overset{\backsim }{H}_{%
\overline{A}_{1}}\left( \overset{\backsim }{\gamma }\right) =g^{-1}\overset{%
\backsim }{H}_{\overline{A}_{2}}\left( \overset{\backsim }{\gamma }\right) g$
for some $g\in G$ and all $\overset{\backsim }{\gamma }\in S^{\ast }$.
Denoting by $\pi \left( S^{\ast }\right) $ the projection on the quotient
space $\left( \overline{\mathcal{A}/\mathcal{G}}\right) /\backsim $,
cylindrical functions are the pull-backs under $\pi \left( S^{\ast }\right) $
of the functions $f$ on $\left( \overline{\mathcal{A}/\mathcal{G}}\right)
/\sim $. $\left( \overline{\mathcal{A}/\mathcal{G}}\right) /\sim $ is
isomorphic to $G^{n}/Ad$ , the algebra of the cylindrical functions is a $%
C^{\ast }$ algebra and its completion in the sup norm is isomorphic to the
algebra $\overline{\mathcal{HA}}$.

A natural integration measure for the cylindrical functions is the Haar
measure on $G$, which being invariant under $G$, projects down naturally to $%
G^{n}/Ad$ . It satisfies the required compatibility condition in the sense
that if $f$ is a cylindrical function on $\overline{\mathcal{A}/\mathcal{G}}$
with respect to two different finitely generated hoop subgroups $S_{i}^{\ast
},i=1,2$, then 
\begin{equation*}
\int_{G^{n_{1}}/Ad}f_{1}d\mu _{1}=\int_{G^{n_{2}}/Ad}f_{2}d\mu _{2}
\end{equation*}%
The measure on $\overline{\mathcal{A}/\mathcal{G}}$ whose restriction to
cylindrical functions is the Haar measure on $G^{n}/Ad$ is the
Ashtekar-Lewandowski measure.

\subsection{Projective limits and physical interpretation}

Let $\left\{ I,\succ \right\} $ be a set endowed with an order relation $%
\succ $ and suppose that with each element $i\in I$ a set $X_{i}$ is
associated and for each pair $\left( i,j\right) $, in which $j\succ i$,
there is a mapping $\pi _{ij}:X_{j}\rightarrow X_{i}$ such that $\pi _{ii}$
is the identity and $\pi _{ki}\pi _{ij}=\pi _{kj}$. Then a set $X$ is called
the projective limit$\mathcal{\ }X=\underset{\longleftarrow }{\lim }X_{i}$
of the family $\left\{ X_{i}\right\} $ of sets if the following conditions
are satisfied:

a) there is a family of mappings $\pi _{i}:X\rightarrow X_{i}$ such that for
any pair $\left( i,j\right) $, in which $j\succ i$, $\pi _{ij}\pi _{j}=\pi
_{i}$

b) for any family of mappings $\alpha _{i}:Y\rightarrow X_{i}$, from an
arbitrary set $Y$, for which the equalities $\pi _{ij}\alpha _{j}=\alpha
_{i} $ hold for $j\succ i$, there exists a unique mapping $\alpha
:Y\rightarrow X$ such that $\alpha _{i}=\pi _{i}\alpha $ for every $i\in I$.

An explicit construction of the projective limit, which is particularly
suited to the physical interpretation is the following: Consider the direct
product $\underset{i\in I}{\prod }X_{i}$ and select in it the subset $X$
which satisfies the consistency condition $j\succ i\Longrightarrow \pi _{ij}%
\overset{\backsim }{X}_{j}=\overset{\backsim }{X}_{i}$ with $\overset{%
\backsim }{X}_{i}\subset X_{i}$ and $\overset{\backsim }{X}_{j}\subset X_{j}$%
. This subset is the projective limit of the family $\left\{ X_{i}\right\} $
of sets.

Notice that this construction emphasizes the fact that the projective limit
is not the limit of the sequence $X_{i}$. Instead it is a particular subset
of the direct product. If, for example, the physical meaning of the index
set $\left\{ I,\succ \right\} $ is a refinement to successive smaller
scales, the projective limit, once the consistency condition is fulfilled,
contains a description of all the scales and not only the small scale limit.

\section{The gauge projective space, kinematical measure, functions and
operators}

In the past, the Ashtekhar-Lewandowski measure has been constructed in very
general settings, using projective limits of floating lattices and weak
smoothness conditions. Here, staying closer to the usual physical setting of
lattice gauge theory, one uses fixed square lattices with piecewise analytic
parametrization.

Consider a sequence of square lattices in $\mathbb{R}^{4}$ of edge $\frac{a}{%
2^{k}}$ $k=0,1,2,\cdots $ constructed in such a way that the lattice of edge 
$\frac{a}{2^{k}}$ is a refinement of the $\frac{a}{2^{k-1}}$ lattice (all
vertices of the $\frac{a}{2^{k-1}}$ lattice are also vertices in the $\frac{a%
}{2^{k}}$ lattice). Finite volume hypercubes $\Gamma $ in this lattice are a
directed set $\left\{ \Gamma ,\succ \right\} $ under the inclusion relation $%
\succ $. $\Gamma \succ \Gamma ^{\prime }$ meaning that all edges and
vertices in $\Gamma ^{\prime }$ are contained in $\Gamma $, the inclusion
relation satisfies 
\begin{eqnarray}
\Gamma &\succ &\Gamma  \notag \\
\Gamma &\succ &\Gamma ^{\prime }\text{ and }\Gamma ^{\prime }\succ \Gamma
\Longrightarrow \Gamma =\Gamma ^{\prime }  \notag \\
\Gamma &\succ &\Gamma ^{\prime }\text{ and }\Gamma ^{\prime }\succ \Gamma
^{\prime \prime }\Longrightarrow \Gamma \succ \Gamma ^{\prime \prime }
\label{3.1}
\end{eqnarray}%
For convenience one considers that the lattice refinement from size $\frac{a%
}{2^{k-1}}$ to $\frac{a}{2^{k}}$ is made one plaquette at a time so that all
intermediate configurations are present in the directed set. This directed
set will cover both successively higher volumes and finer and finer
lattices. Let $p_{0}$ be a point that does not belong to any lattice of the
directed family. One assumes an analytic parametrization of each edge, to
each edge $l$ associates a $p_{0}$-based loop and for each generalized
connection $A$ consider the holonomy $h_{l}\left( A\right) $.

For definiteness each edge is considered to be oriented along the
coordinates positive direction and the set of edges of the lattice $\Gamma $
is denoted $E\left( \Gamma \right) $. The set $\mathcal{A}_{\Gamma }$ of
generalized connections for the lattice hypercube $\Gamma $ is the set of
homorphisms $\mathcal{A}_{\Gamma }=Hom\left( E\left( \Gamma \right)
,G\right) \sim G^{\#E\left( \Gamma \right) }$, obtained by associating to
each edge the holonomy $h_{l}\left( \cdot \right) $ on the associated $p_{0}$%
-based loop. The set of gauge-independent generalized connections $\mathcal{A%
}_{\Gamma }/Ad$ is obtained factoring by the adjoint representation at $%
p_{0} $, $\mathcal{A}_{\Gamma }/Ad$ $\sim G^{\#E\left( \Gamma \right) }/Ad$.
However because, for gauge independent functions, integration in $\mathcal{A}%
_{\Gamma }$ coincides with integration in $\mathcal{A}_{\Gamma }/Ad$ , for
simplicity from now on one uses only $\mathcal{A}_{\Gamma }$. The space of
generalized connections that one considers here is then the projective limit 
$\mathcal{A}=\underset{\longleftarrow }{\lim }\mathcal{A}_{\Gamma }$ of the
family 
\begin{equation}
\left\{ \mathcal{A}_{\Gamma },\pi _{\Gamma \Gamma ^{\prime }}:\Gamma
^{\prime }\succ \Gamma \right\}  \label{3.2}
\end{equation}%
$\pi _{\Gamma \Gamma ^{\prime }}$ and $\pi _{\Gamma }$ denote the surjective
projections $\mathcal{A}_{\Gamma ^{\prime }}\longrightarrow \mathcal{A}%
_{\Gamma }$ and $\mathcal{A}=\underset{\longleftarrow }{\lim }\mathcal{A}%
_{\Gamma }$.

Recall that the projective limit of the family $\left\{ \mathcal{A}_{\Gamma
},\pi _{\Gamma \Gamma ^{\prime }}\right\} $ is the subset $\mathcal{A}$ of
the Cartesian product $\underset{\Gamma }{\prod }\mathcal{A}_{\Gamma }$ that
satisfies the consistent condition 
\begin{equation*}
\mathcal{A}=\left\{ a\in \underset{\Gamma }{\prod }\mathcal{A}_{\Gamma
}:\Gamma ^{\prime }\succ \Gamma \Longrightarrow \pi _{\Gamma \Gamma ^{\prime
}}\mathcal{A}_{\Gamma ^{^{\prime }}}=\mathcal{A}_{\Gamma }\right\}
\end{equation*}
The projective topology in $\mathcal{A}$ is the coarsest topology for which
each $\pi _{\Gamma }$ mapping is continuous.

For a compact group $G$, each $\mathcal{A}_{\Gamma }$ is a compact Hausdorff
space. Then $\mathcal{A}$ is also a compact Hausdorff space. In each $%
\mathcal{A}_{\Gamma }$ one has a natural (Haar) normalized product measure $%
\nu _{\Gamma }=\mu _{H}^{\#E\left( \Gamma \right) }$, $\mu _{H}$ being the
normalized Haar measure in $G$. Then, according to a theorem of Prokhorov,
as generalized by Kisynski\cite{Kisynski} \cite{Maurin}, if 
\begin{equation}
\nu _{\Gamma ^{\prime }}\left( \pi _{\Gamma \Gamma ^{\prime }}^{-1}\left(
B\right) \right) =\nu _{\Gamma }\left( B\right)  \label{3.2a}
\end{equation}%
for every $\Gamma ^{\prime }\succ \Gamma $ and every Borel set $B$ in $%
\mathcal{A}_{\Gamma }$, there is a unique measure $\nu $ in $\mathcal{A}$
such that $\nu \left( \pi _{\Gamma }^{-1}\left( B\right) \right) =\nu
_{\Gamma }\left( B\right) $ for every $\Gamma $. Furthermore, this measure
is tight, that is, for every $\varepsilon >0$ there is a compact subset $K$
of $\mathcal{A}$ such that $\nu _{\Gamma }\left( \mathcal{A}_{\Gamma }-\pi
_{\Gamma }\left( K\right) \right) <\varepsilon $. The measure $\nu $, so
constructed, is a version of the Ashtekar-Lewandowski measure.

The consistency condition (\ref{3.2a}) is easy to check in the present
context. It suffices to consider $\Gamma ^{\prime }=\Gamma _{k}$ as the
refinement of $\Gamma =\Gamma _{k-1}$ when the edge size goes from $\frac{a}{%
2^{k-1}}$ to $\frac{a}{2^{k}}$. Then, if $g_{i}$ are group elements
associated to the finer lattice (size $\frac{a}{2^{k}}$) 
\begin{eqnarray}
\nu _{k-1}\left( g_{1}\times g_{2}\times g_{3}\times g_{4}\in B\right)
&=&\nu _{k-1}\left( g_{i}\times g_{j}\times g_{k}\times \left( g_{i}\times
g_{j}\times g_{k}\right) ^{-1}B:g_{i},g_{j},g_{k}\in G\right)  \notag \\
&=&\mu _{H}\left( G\right) ^{3}\mu _{H}\left( \left( g_{i}\times g_{j}\times
g_{k}\right) ^{-1}B\right) =\nu _{k}\left( B\right)  \label{3.2b}
\end{eqnarray}%
and the consistency condition (\ref{3.2a}) follows from the normalization
and invariance of the $\mu _{H}$ measure. The second equality in (\ref{3.2b}%
) reflects the factorized nature of the product measure. $\mathcal{A}=%
\underset{\longleftarrow }{\lim }\mathcal{A}_{\Gamma }$ will be called 
\textit{the gauge space} and $\nu $ \textit{the kinematical measure}.

In the following one also needs to define functions and operators in the
projective family. The correspondence $\mathcal{A}_{\Gamma }\thicksim
G^{\#E\left( \Gamma \right) }$ means that functions on the projective family
are constructed from equivalent classes of functions in $G^{\#E\left( \Gamma
\right) }$.

$G$ being a compact connected Lie group with Lie algebra $LG$, one chooses
an $AdG-$invariant inner product $\left( \cdot ,\cdot \right) $ on $LG$. For
each $\xi \in LG$ define $\partial _{\xi }$ by 
\begin{equation}
\left( \partial _{\xi }f\right) \left( g\right) :=\left. \frac{d}{dt}f\left(
e^{t\xi }g\right) \right\vert _{t=0};\qquad g\in G  \label{3.3}
\end{equation}%
Choosing an orthonormal basis in $LG$, $\left\{ \xi _{1}\cdots \xi
_{n}\right\} $, write $\partial _{i}:=\partial _{\xi _{i}}$. With these
operators one has a notion of $C^{n}\left( G\right) $ functions and, with
the Haar measure $d\mu _{H}$, of $L^{2}\left( G,d\mu _{H}\right) $ space as
well.

The Laplacian operator is 
\begin{equation}
\Delta _{G}:=\sum_{i=1}^{n}\partial _{i}^{2}  \label{3.4}
\end{equation}
which does not depend on the choice of the basis and is symmetric with
respect to the $L^{2}\left( G,d\mu _{H}\right) $ inner product.

For any finite $\#E\left( \Gamma \right) $, the extension of these notions
to $\mathcal{A}_{\Gamma }\thicksim G^{\#E\left( \Gamma \right) }$ is
straightforward. To carry the notion of $C^{n}$ function in (finite) product
spaces to the projective family, introduce in the union 
\begin{equation*}
\underset{\Gamma }{\bigcup }C^{n}\left( \mathcal{A}_{\Gamma }\right)
\end{equation*}
the equivalence relation 
\begin{equation}
f_{\Gamma _{1}}\thicksim f_{\Gamma _{2}}\qquad \text{if}\qquad \pi _{\Gamma
_{1}\Gamma _{3}}^{*}f_{\Gamma _{1}}=\pi _{\Gamma _{2}\Gamma
_{3}}^{*}f_{\Gamma _{2}}  \label{3.5}
\end{equation}
for any $\Gamma _{3}\succ \Gamma _{1},\Gamma _{2}$. $\pi _{\Gamma \Gamma
^{^{\prime }}}^{*}$ is the pull-back map from the space of functions on $%
\Gamma $ to the space of functions on $\Gamma ^{^{\prime }}$.

The set of $C^{n}$ cylindrical functions associated to the projective family 
$\left\{ \mathcal{A}_{\Gamma },\pi _{\Gamma \Gamma ^{\prime }}\right\} $ is
then 
\begin{equation}
\underset{\Gamma }{\bigcup }C^{n}\left( \mathcal{A}_{\Gamma }\right)
/\thicksim  \label{3.6}
\end{equation}

On the other hand, for families of operators $\left\{ O_{\Gamma },D\left(
O_{\Gamma }\right) \right\} _{\Gamma \in S}$ with domains $D\left( O_{\Gamma
}\right) $ defined on a subset of labels $S$, one requires the following
consistency conditions 
\begin{equation}
\pi _{\Gamma \Gamma ^{^{\prime }}}^{*}D\left( O_{\Gamma }\right) \subset
D\left( O_{\Gamma ^{^{\prime }}}\right)  \label{3.7a}
\end{equation}
\begin{equation}
O_{\Gamma ^{^{\prime }}}\pi _{\Gamma \Gamma ^{^{\prime }}}^{*}=\pi _{\Gamma
\Gamma ^{^{\prime }}}^{*}O_{\Gamma }  \label{3.7b}
\end{equation}
for every $\Gamma ^{^{\prime }}\succ \Gamma $ such that $\Gamma ,\Gamma
^{^{\prime }}\in S$.

\section{The gauge projective triplet}

Here one considers the same directed set $\left\{ \Gamma ,\succ \right\} $,
spaces $\mathcal{A}_{\Gamma }=Hom\left( E\left( \Gamma \right) ,G\right)
\sim G^{\#E\left( \Gamma \right) }$ and the projective limit $\mathcal{A}=%
\underset{\longleftarrow }{\lim }\mathcal{A}_{\Gamma }$ as in the preceding
section. To formulate an infinite-dimensional calculus in $\mathcal{A}$ one
starts by defining test functions and distributions in $G^{\#E\left( \Gamma
\right) }$ and then considers the corresponding projective limits.

Several families of norms may be used to construct test functions and
distribution spaces in $G^{\#E\left( \Gamma \right) }$. Here the heat kernel
norm will be used. For a compact connected group $G$, one uses the same
construction of Hida spaces as in Ref.\cite{Deck} to which the reader is
referred for details and proofs. In $G$ there is a heat kernel $p_{t}\left(
g\right) $ defined as the fundamental solution of 
\begin{equation}
\partial _{t}p=\frac{1}{2}\Delta p  \label{4.1}
\end{equation}%
$\Delta $ being the operator defined in (\ref{3.4}). The heat kernel measure
is $d\mu _{t}=p_{t}d\mu _{H}$ and it is easy to check that 
\begin{equation}
L^{2}\left( G,d\mu _{t}\right) =L^{2}\left( G,d\mu _{H}\right)  \label{4.2}
\end{equation}

By analogy with the Gaussian case a collection $\mathcal{H}_{t}\left(
G\right) $ of spaces is defined as domains of 
\begin{equation}
\mathcal{H}_{t}\left( G\right) =\mathcal{D}\left( p_{1/t}^{-1/2}\circ e^{-%
\frac{1}{2}\left( 1-\frac{1}{t}\right) \bigtriangleup }\circ p_{1}\right)
\label{4.3}
\end{equation}%
for $t\geq 1$, which are Hilbert spaces with norm 
\begin{equation}
\left\Vert f\right\Vert _{t}^{2}=\int_{G}\left\vert \left(
p_{1/t}^{-1/2}\circ e^{-\frac{1}{2}\left( 1-\frac{1}{t}\right)
\bigtriangleup }\circ p_{1}\right) f\right\vert ^{2}d\mu _{H}  \label{4.4}
\end{equation}%
Equivalently, if the eigenvalues of the Laplacian are $\varphi _{n}$%
\begin{equation}
\bigtriangleup \varphi _{n}=\lambda _{n}\varphi _{n}  \label{4.5}
\end{equation}%
then 
\begin{equation}
f\in \mathcal{H}_{t}\left( G\right) \Longleftrightarrow f=\sum_{n=1}^{\infty
}c_{n}\frac{\varphi _{n}}{p_{1}}  \label{4.6}
\end{equation}%
with 
\begin{equation}
\sum_{n=1}^{\infty }\left\vert c_{n}\right\vert ^{2}e^{-\left( 1-\frac{1}{t}%
\right) \lambda _{n}}<\infty  \label{4.7}
\end{equation}%
Because $\bigtriangleup $ has negative spectrum, $\mathcal{H}_{t^{\prime
}}\left( G\right) \subset \mathcal{H}_{t}\left( G\right) $ if $t^{\prime }>t$

From the family $\left\{ \mathcal{H}_{t}\left( G\right) ,t\geq 1\right\} $
of Hilbert spaces one defines the \textit{test function space on} $G$ as 
\begin{equation}
\mathcal{H}\left( G\right) =\underset{t\geq 1}{\bigcap }\mathcal{H}%
_{t}\left( G\right) =\underset{n\in \mathbb{N}}{\bigcap }\mathcal{H}%
_{n}\left( G\right)  \label{4.8}
\end{equation}%
$\mathcal{H}\left( G\right) $ is equipped with the projective limit topology
of the spaces $\mathcal{H}_{t}\left( G\right) $, which coincides with the
metric topology defined by the metric 
\begin{equation}
d\left( f,g\right) =\sum_{n=1}^{\infty }\frac{1}{2^{n}}\frac{\left\Vert
f-g\right\Vert _{n}}{1+\left\Vert f-g\right\Vert _{n}}  \label{4.9}
\end{equation}%
$\mathcal{H}\left( G\right) $ is dense in each $\mathcal{H}_{t}\left(
G\right) $ and is a nuclear space of analytic functions on $G$ \cite{Deck}.

Because $\left( \mathcal{H}\left( G\right) ,d\right) $ is a countably
Hilbert space it follows \cite{Gelfand} that the topological dual $\mathcal{H%
}^{\ast }\left( G\right) $ of $\mathcal{H}\left( G\right) $ is given by 
\begin{equation}
\mathcal{H}^{\ast }\left( G\right) =\bigcup_{n=1}^{\infty }\mathcal{H}%
_{n}^{\ast }\left( G\right)  \label{4.10}
\end{equation}%
$\mathcal{H}_{n}^{\ast }\left( G\right) $ being the dual space of $\mathcal{H%
}_{n}^{\ast }\left( G\right) $. That is, each continuous linear functional
on $\mathcal{H}\left( G\right) $ must already be continuous for some norm $%
\left\Vert \bullet \right\Vert _{n}$. The nuclearity of $\mathcal{H}\left(
G\right) $ also implies that $\mathcal{H}^{\ast }\left( G\right) $ carries
many probability measures defined by characteristic functions and the
Bochner-Minlos theorem. $\mathcal{H}^{\ast }\left( G\right) $ is the \textit{%
space of distributions} on $G$. By a canonical embedding one has the chain
(triplet) 
\begin{equation}
\mathcal{H}\left( G\right) \subset L^{2}\left( G,d\mu _{t}\right) \subset 
\mathcal{H}^{\ast }\left( G\right)  \label{4.11}
\end{equation}

For each finite hypercube $\Gamma $, taking direct products of $\#E\left(
\Gamma \right) $ copies of the spaces, the generalization of this triplet
construction to $G^{\#E\left( \Gamma \right) }$is straightforward 
\begin{equation}
\mathcal{H}\left( G^{\#E\left( \Gamma \right) }\right) =\underset{t\geq 1}{%
\bigcap }\mathcal{H}_{t}\left( G^{\#E\left( \Gamma \right) }\right) =%
\underset{n\in \mathbb{N}}{\bigcap }\mathcal{H}_{n}\left( G^{\#E\left(
\Gamma \right) }\right)  \label{4.12}
\end{equation}%
\begin{equation}
\mathcal{H}^{\ast }\left( G^{\#E\left( \Gamma \right) }\right)
=\bigcup_{n=1}^{\infty }\mathcal{H}_{n}^{\ast }\left( G^{\#E\left( \Gamma
\right) }\right)  \label{4.13}
\end{equation}%
\begin{equation}
\mathcal{H}\left( G^{\#E\left( \Gamma \right) }\right) \subset L^{2}\left(
G^{\#E\left( \Gamma \right) },\prod_{E\left( \Gamma \right) }d\mu
_{t}\right) \subset \mathcal{H}^{\ast }\left( G^{\#E\left( \Gamma \right)
}\right)  \label{4.14}
\end{equation}%
This provides, for each finite hypercube $\Gamma $, a space of test
functions $\mathcal{H}\left( G^{\#E\left( \Gamma \right) }\right) $ and
distributions $\mathcal{H}^{\ast }\left( G^{\#E\left( \Gamma \right)
}\right) $ on $\mathcal{A}_{\Gamma }=Hom\left( E\left( \Gamma \right)
,G\right) \sim G^{\#E\left( \Gamma \right) }$.

With the directed set $\left\{ \Gamma ,\succ \right\} $ of finite volume
hypercubes, one has the surjective projections $\mathcal{H}\left( \mathcal{A}%
_{\Gamma ^{\prime }}\sim G^{\#E\left( \Gamma ^{\prime }\right) }\right) 
\overset{\pi _{\Gamma \Gamma ^{\prime }}^{S}}{\longrightarrow }\mathcal{H}%
\left( \mathcal{A}_{\Gamma }\sim G^{\#E\left( \Gamma \right) }\right) $ and

$\mathcal{H}^{\ast }\left( \mathcal{A}_{\Gamma ^{\prime }}\sim G^{\#E\left(
\Gamma ^{\prime }\right) }\right) \overset{\pi _{\Gamma \Gamma ^{\prime
}}^{S^{\ast }}}{\longrightarrow }\mathcal{H}^{\ast }\left( \mathcal{A}%
_{\Gamma }\sim G^{\#E\left( \Gamma \right) }\right) $ for $\Gamma ^{\prime
}\succ \Gamma $, the maps $\pi _{\Gamma \Gamma ^{\prime }}^{S}$ and $\pi
_{\Gamma \Gamma ^{\prime }}^{S^{\ast }}$ meaning the restriction of
functions and distributions on $G^{\#E\left( \Gamma ^{\prime }\right) }$ to
the elements of $G^{\#E\left( \Gamma \right) }$.

One now considers, in the Cartesian products $\underset{\Gamma }{\prod }%
\mathcal{H}\left( \mathcal{A}_{\Gamma }\right) $ and $\underset{\Gamma }{%
\prod }\mathcal{H}^{\ast }\left( \mathcal{A}_{\Gamma }\right) $ the subsets 
\begin{equation}
\mathcal{H}\left( \mathcal{A}\right) =\left\{ s\left( a\right) \in \underset{%
\Gamma }{\prod }\mathcal{H}\left( \mathcal{A}_{\Gamma }\right) :\Gamma
^{\prime }\succ \Gamma \Longrightarrow \pi _{\Gamma \Gamma ^{\prime }}^{S}%
\mathcal{H}\left( \mathcal{A}_{\Gamma ^{^{\prime }}}\right) =\mathcal{H}%
\left( \mathcal{A}_{\Gamma }\right) \right\}  \label{4.15}
\end{equation}%
and 
\begin{equation}
\mathcal{H}^{\ast }\left( \mathcal{A}\right) =\left\{ s\left( a^{\ast
}\right) \in \underset{\Gamma }{\prod }\mathcal{H}^{\ast }\left( \mathcal{A}%
_{\Gamma }\right) :\Gamma ^{\prime }\succ \Gamma \Longrightarrow \pi
_{\Gamma \Gamma ^{\prime }}^{S^{\ast }}\mathcal{H}^{\ast }\left( \mathcal{A}%
_{\Gamma ^{^{\prime }}}\right) =\mathcal{H}^{\ast }\left( \mathcal{A}%
_{\Gamma }\right) \right\}  \label{4.16}
\end{equation}%
which define spaces of test functions and distributions on $\mathcal{A}$. It
is this projective triplet 
\begin{equation}
\mathcal{H}\left( \mathcal{A}\right) \subset L^{2}\left( \mathcal{A},d\nu
\right) \mathcal{\subset H}^{\ast }\left( \mathcal{A}\right)  \label{4.17}
\end{equation}%
that provides the framework for an infinite-dimensional calculus in the
gauge theory. A particularly useful tool for this purpose is the $S-$%
transform, which for $\mathcal{H}^{\ast }\left( G\right) $ is 
\begin{equation}
\left( S\Phi \right) \left( x\right) =\Phi \left( e_{x}\right)  \label{4.17a}
\end{equation}%
$\Phi \in $ $\mathcal{H}^{\ast }\left( G\right) $, $x\in G$ and 
\begin{equation}
e_{x}\left( y\right) =\frac{p_{1}\left( x^{-1}y\right) }{p_{1}\left(
y\right) }  \label{4.17b}
\end{equation}%
The $S-$transform is an injective map from $\mathcal{H}^{\ast }\left(
G\right) $ onto $\mathcal{U}\left( G_{\mathbb{C}}\right) $, the space of
holomorphic functions of second order exponential growth on $G_{\mathbb{C}}$
(the complexification of $G)$.%
\begin{equation}
\mathcal{U}\left( G_{\mathbb{C}}\right) =\left\{ f\in Hol\left( G_{\mathbb{C}%
}\right) |\exists k,c>0:\left\vert f\left( z\right) \right\vert \leq
ce^{k\left\vert z\right\vert ^{2}},\forall z\in G_{\mathbb{C}}\right\}
\label{4.18}
\end{equation}%
The extension of this transform to $G^{\#E\left( \Gamma \right) }$ is
straightforward and through the Cartesian product construction allows to
deal with distributions in $\mathcal{H}^{\ast }\left( \mathcal{A}\right) $
as functions in $\mathcal{H}\left( \mathcal{A}\right) $.

Notice that all spaces in the gauge projective triplet (\ref{4.17}) are
subsets of a Cartesian product, not just the corresponding small distance
limit. Therefore the triplet, here proposed, is the basic framework for a
gauge theory calculus at all length scales.

The factorizable nature of the $\nu $ measure in $L^{2}\left( \mathcal{A}%
,d\nu \right) $ played an important role in checking the consistency
condition (\ref{3.2a}). However, being factorizable, it is necessarily a
non-interacting measure. Next, one discusses how to define a class of
interacting measures in the gauge triplet framework.

\section{Convolution semigroups and interaction measures}

In (\ref{3.2b}) the consistency condition (\ref{3.2a}) is easy to check
because of the factorized nature of the kinematical measure $\nu $. However,
interaction measures have to be constructed from entities involving more
than one of the edge-based holonomies. The basic element will be 
\begin{equation}
U_{\square }\left( A_{\Gamma }\right) =h_{1}h_{2}h_{3}^{-1}h_{4}^{-1}
\label{5.1}
\end{equation}
$h_{1}$ to $h_{4}$ being the holonomies associated to the loops based on the
links of a plaquette. Then $U_{\square }\left( A_{\Gamma }\right) $ is the
holonomy along the plaquette which, according to the orientation conventions
used here, is obtained by the product of two $p_{0}-$based loop holonomies
and two inverse $p_{0}-$based loop holonomies.

To construct an interaction measure, one first considers, on the
finite-dimensional spaces $\mathcal{A}_{\Gamma }\sim G^{\#E\left( \Gamma
\right) }$, measures that are absolutely continuous with respect to the Haar
measure 
\begin{equation}
d\mu _{\mathcal{A}_{\Gamma }}=\frac{1}{Z}p\left( \mathcal{A}_{\Gamma
}\right) \left( d\mu _{H}\right) ^{\#E\left( \Gamma \right) }  \label{5.2}
\end{equation}
where $p\left( \mathcal{A}_{\Gamma }\right) $ is a continuous function in $%
\mathcal{A}_{\Gamma }$ and a $Z$ a normalizing constant. In particular make
the simplifying assumptions:

- that $p\left( \mathcal{A}_{\Gamma }\right) $ is a product of plaquette
functions 
\begin{equation}
p\left( \mathcal{A}_{\Gamma }\right) =p\left( U_{\square _{1}}\right)
p\left( U_{\square _{2}}\right) \cdots p\left( U_{\square _{n}}\right)
\label{5.3}
\end{equation}

the product running over the $n$ plaquettes contained in $\Gamma $ and

- that $p\left( \cdot \right) $ is a central function, $p\left( xy\right)%
=p\left( yx\right) $ or, equivalently $p\left( y^{-1}xy\right) =p\left(%
x\right)$ with $x,y \in G$.

To be able to construct an interaction measure on the projective limit one
has to check the consistency condition (\ref{3.2a}). In the directed set $%
\left\{ \Gamma ,\succ \right\} $ consider two elements $\Gamma $ and $\Gamma
^{^{\prime }}$ which differ only in subdivision of a single plaquette (see
Fig.1), all the others being the same.

\begin{figure}[tbh]
\begin{center}
\psfig{figure=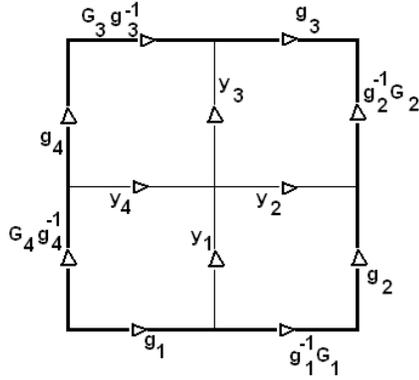,width=11truecm}
\end{center}
\caption{Subdivision of one plaquette}
\end{figure}

If there is a choice of densities $p\left( \cdot \right) $ that fulfill the
consistency condition in this case, then it can be satisfied for whole
directed set. For this case the consistency condition is simply 
\begin{eqnarray}
&&\frac{1}{Z^{^{\prime }}}\int p^{\prime }\left(
g_{1}^{-1}G_{1}g_{2}y_{2}^{-1}y_{1}^{-1}\right) p^{\prime }\left(
y_{2}g_{2}^{-1}G_{2}g_{3}^{-1}y_{3}^{-1}\right) p^{\prime }\left(
y_{4}y_{3}g_{3}G_{3}^{-1}g_{4}^{-1}\right) p^{\prime }\left(
g_{1}y_{1}y_{4}^{-1}g_{4}G_{4}^{-1}\right)  \notag \\
&&\prod_{i=1}^{4}d\mu _{H}\left( g_{i}\right) d\mu _{H}\left( y_{i}\right)
d\mu _{H}\left( G_{i}\right)  \notag \\
&=&\frac{1}{Z}\int p\left( G_{1}G_{2}G_{3}^{-1}G_{4}^{-1}\right)
\prod_{i=1}^{4}d\mu _{H}\left( G_{i}\right)  \label{5.4}
\end{eqnarray}%
because integration over all the other plaquettes is the same in $\Gamma $
and $\Gamma ^{^{\prime }}$. $p^{\prime }$ and $p$ denote the densities for
plaquettes of size $\frac{a}{2^{k}}$ and $\frac{a}{2^{k-1}}$, respectively.

Using centrality of $p^{\prime }$, redefining 
\begin{equation}
g_{1}y_{1}=X_{1},\qquad g_{2}y_{2}^{-1}=X_{2},\qquad
y_{3}g_{3}=X_{3}^{-1},\qquad y_{4}^{-1}g_{4}=X_{4}^{-1}  \label{5.5}
\end{equation}%
and using invariance of the normalized Haar measure, one may integrate over $%
y_{1},y_{2},y_{3},y_{4}$ obtaining for the left hand side of (\ref{5.4}) 
\begin{equation*}
\frac{1}{Z^{^{\prime }}}\int p^{\prime }\left( X_{1}^{-1}G_{1}X_{2}\right)
p^{\prime }\left( X_{2}^{-1}G_{2}X_{3}\right) p^{\prime }\left(
X_{3}^{-1}G_{3}^{-1}X_{4}\right) p^{\prime }\left(
X_{4}^{-1}G_{4}^{-1}X_{1}\right) \prod_{i=1}^{4}d\mu _{H}\left( X_{i}\right)
d\mu _{H}\left( G_{i}\right)
\end{equation*}%
Finally, if there is a sequence of central functions $p^{\prime },p^{\prime
\prime },p$ satisfying 
\begin{eqnarray}
\int p^{\prime }\left( G_{i}X\right) p^{\prime }\left( X^{-1}G_{j}\right)
d\mu _{H}\left( X\right) &\sim &p^{\prime \prime }\left( G_{i}G_{j}\right) 
\notag \\
\int p^{\prime \prime }\left( G_{i}X\right) p^{\prime \prime }\left(
X^{-1}G_{j}\right) d\mu _{H}\left( X\right) &\sim &p\left( G_{i}G_{j}\right)
\label{5.6}
\end{eqnarray}%
the consistency condition (\ref{5.4}) would be satisfied, the
proportionality constants being absorbed by the normalization constant $%
Z^{^{\prime }}$. $p^{\prime },p^{\prime \prime }$ and $p$ are the functions
associated to the square plaquette with links of size $\frac{a}{2^{k}}$, the
rectangular plaquette with links of size $\frac{a}{2^{k}}$ and $\frac{a}{%
2^{k-1}}$ and, finally, the square plaquette with links of size $\frac{a}{%
2^{k-1}}$. The sequence $\left( p^{\prime },p^{\prime \prime },p\right) $
corresponds to the subdivision of one plaquette. If such a sequence exists
for all $k$, because all elements in the directed set $\left\{ \Gamma ,\succ
\right\} $ may be reached by one-plaquette subdivisions, one obtains the
following general result:

\begin{theorem}
\textit{An interaction measure on the projective limit }$\mathcal{A}=%
\underset{\longleftarrow }{\lim }\mathcal{A}_{\Gamma }$\textit{\ exists if
a sequence of functions is found satisfying (\ref{5.6}) for plaquette 
subdivisions of all sizes.}
\end{theorem}

Notice that:

- Eq.(\ref{5.6}) is not necessarily a convolution semigroup property because 
$p^{\prime },p^{\prime \prime }$ and $p$ might be different functions and (%
\ref{5.6}) is a proportionality relation, not an equality.

- The interaction measure may not be absolutely continuous with respect to
the kinematical measure $\nu $, constructed in Sect.2, because not all
functions in the sequence (mostly in the small scale limit) might be
continuous functions.

Although (\ref{5.6}) is not exactly a convolution semigroup property,
functions satisfying this condition may be obtained out of convolution
semigroup kernels. Three cases will be separately analyzed, namely $%
G=U\left( 1\right) ,SU\left( 2\right) ,SU\left( 3\right) $.

\subsection{$G=U\left( 1\right) $}

An important convolution semigroup in $U\left( 1\right) $ 
\begin{equation}
U\left( 1\right) =\left\{ e^{i2\pi \theta };\theta \in \left[ 0,1\right)
\right\}  \label{5.7}
\end{equation}%
is the heat kernel semigroup 
\begin{equation}
K_{1}\left( e^{i2\pi \theta },\beta \right) =\sum_{n\in \mathbb{Z}}\exp
\left\{ i2\pi n\theta -\left( 2\pi n\right) ^{2}\beta \right\}  \label{5.8}
\end{equation}%
which, by convolution with any initial condition $u_{0}\left( \theta \right) 
$, provides a solution to the $U\left( 1\right) -$heat equation 
\begin{equation}
\left( \frac{\partial }{\partial \beta }-\frac{\partial ^{2}}{\partial
\theta ^{2}}\right) u\left( e^{i2\pi \theta },\beta \right) =0  \label{5.9}
\end{equation}%
Let us now use the $U\left( 1\right) -$heat kernel in (\ref{5.8}) with 
\begin{equation}
p\left( e^{i2\pi \theta }\right) =K_{1}\left( e^{i2\pi \theta },\beta \right)
\label{5.10}
\end{equation}%
to check the condition (\ref{5.6}). One obtains 
\begin{equation}
\int K_{1}\left( e^{i2\pi \left( \theta _{1}+\alpha \right) },\beta ^{\prime
}\right) K_{1}\left( e^{i2\pi \left( \theta _{2}-\alpha \right) },\beta
^{\prime }\right) d\alpha =K_{1}\left( e^{i2\pi \left( \theta _{1}+\theta
_{2}\right) },2\beta ^{\prime }\right)  \label{5.11}
\end{equation}%
Iterating this relation, one concludes that the measure consistency
condition is satisfied with the choice (\ref{5.10}) if 
\begin{equation}
\beta ^{\prime }=\frac{\beta }{4}  \label{5.12}
\end{equation}%
that is, each time one plaquette is subdivided the \textquotedblleft
time\textquotedblright\ label $\beta $ in the densities associated to that
plaquette should be divided by $4$. Therefore, using the heat kernel as the
density of the measure, a consistent measure is constructed in the
projective limit.

An important consideration in establishing measures for the gauge spaces is
to check whether, in the small scales, these measures correspond (or not) to
the measures used by physicists for the same phenomena. In the $U\left(
1\right) $ case this is easily seen by rewriting the heat kernel using the
Jacobi identity 
\begin{equation}
K_{1}\left( e^{i2\pi \theta },\beta \right) =\frac{1}{\sqrt{4\pi \beta }}%
\sum_{n\in \mathbb{Z}}\exp \left\{ -\frac{\left( \theta +n\right) ^{2}}{%
4\beta }\right\}  \label{5.13}
\end{equation}%
Then, at very small scales $\beta $ becomes extremely small. Therefore in
the sum of (\ref{5.13}) only the $n=0$ and the $\theta \approx 0$
neighborhood contribute and, for the plaquette, one obtains a density
proportional to 
\begin{equation}
\exp \left\{ -\frac{\left( \theta _{1}+\theta _{2}-\theta _{3}-\theta
_{4}\right) ^{2}}{4\beta }\right\}  \label{5.14}
\end{equation}%
which indeed corresponds to the usual (small scale) $U\left( 1\right) -$%
measure.

In the lattice used to define the $\left\{ \Gamma ,\succ \right\} $ directed
set when, in the lattice size $\frac{a}{2^{k}}$, $k\rightarrow \infty $ then 
$\beta \rightarrow 0$. In this limit the density (\ref{5.13}) is no longer
continuous, therefore, the interaction measure is not absolutely continuous
with respect to the kinematical $\nu $ measure. Nevertheless the interaction
measure has a generalized density in $\mathcal{H}^{\ast }\left( \mathcal{A}%
\right) $, the distribution space of the projective gauge triplet described
before. Summarizing:

\begin{theorem}
\textit{Using the density (\ref{5.13}) for each link-based loop, with }$%
\beta _{k}=c\frac{1}{4^{k}}$ \textit{\ (}$c$\textit{\ an
arbitrary constant) for each }$\Gamma _{k}$\textit{\ in the directed set, a
consistent measure is shown to exist in the projective limit }$\mathcal{A}=%
\underset{\longleftarrow }{\lim }\mathcal{A}_{\Gamma _{k}}$\textit{\ of
the }$U\left( 1\right) $\textit{\ gauge theory. Furthermore, the measure
coincides at small scales with the Gaussian measure (\ref{5.14}). The
measure in the projective limit is not absolutely continuous with respect to
the kinematical measure, but it has a generalized density in }$\mathcal{H}%
^{*}\left( \mathcal{A}\right) $\textit{.\bigskip }
\end{theorem}

In Fig.2 one plots the density (\ref{5.13}) for $\theta \in \left[ 0,1\right]
$ and $\beta \in \left[ 0.001,0.4\right] $. One sees that for small $\beta $
(small scales) the measure concentrates around $\theta =0$.

\begin{figure}[tbh]
\begin{center}
\psfig{figure=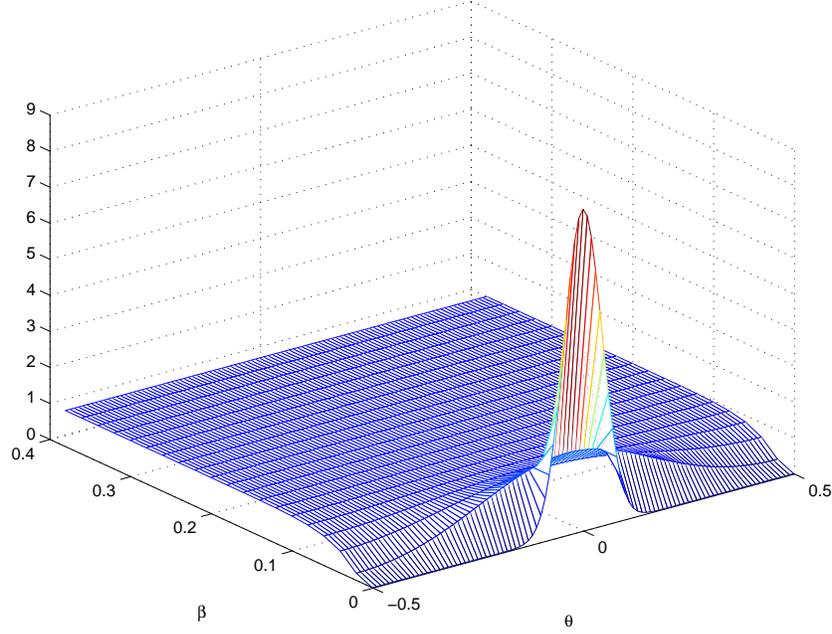,width=11truecm}
\end{center}
\caption{The $U\left( 1\right) -$heat kernel density}
\end{figure}

\subsection{Non-abelian compact groups}

Now that the $U\left( 1\right) $ case is understood, a simple argument shows
that a similar construction is possible for general compact groups. In a
compact Lie group the heat kernel is 
\begin{equation}
K\left( g,\beta \right) =\sum_{\lambda \in \Lambda ^{+}}d_{\lambda
}e^{-c\left( \lambda \right) \beta }\chi _{\lambda }\left( g\right)
\label{5.15}
\end{equation}%
with $g\in G$ and $\beta \in \mathbb{R}^{+}$. $\Lambda ^{+}$ is the set of
highest weights, $d_{\lambda }$ and $\chi _{\lambda }\left( \cdot \right) $
the dimension and the character of the $\lambda -$representation and $%
c\left( \lambda \right) $ the spectrum of the Laplacian (\ref{3.4}) 
\begin{equation}
\left( \Delta _{G}\chi _{\lambda }\right) \left( g\right) =c\left( \lambda
\right) \chi _{\lambda }\left( g\right)  \label{5.16}
\end{equation}%
Using, as before, the heat kernel for the construction of the interaction
measure, the condition (\ref{5.6}) becomes 
\begin{eqnarray}
&&\int d\mu _{H}\left( x\right) K\left( g_{1}x,\beta \right) K\left(
x^{-1}g_{2},\beta \right)  \notag \\
&=&\sum_{\lambda ,\lambda ^{\prime }\in \Lambda ^{+}}d_{\lambda }d_{\lambda
^{\prime }}e^{-c\left( \lambda \right) \beta }e^{-c\left( \lambda ^{\prime
}\right) \beta }\int d\mu _{H}\left( x\right) \chi _{\lambda }\left(
g_{1}x\right) \chi _{\lambda ^{\prime }}\left( x^{-1}g_{2}\right)  \notag \\
&=&K\left( g_{1}g_{2},2\beta \right)  \label{5.17}
\end{eqnarray}%
the last equality following from Schur's orthogonality relations. Therefore,
whenever the heat kernel is chosen as the density for the loops, the
situation is quite similar to the $U\left( 1\right) $ case, namely, on each
subdivision of a plaquette 
\begin{equation}
\beta \rightarrow \beta ^{\prime }=\frac{\beta }{4}  \label{5.18}
\end{equation}%
Hence

\begin{theorem}
\textit{Using the heat kernel (\ref{5.15}) for the density of each plaquette,
with }$\beta _{k}=c\frac{1}{4^{k}}$ \textit{\ (}$c$\textit{\
an arbitrary constant) for each }$\Gamma _{k}$\textit{\ in the directed set,
a consistent measure is shown to exist in the projective limit }$\mathcal{A}=%
\underset{\longleftarrow }{\lim }\mathcal{A}_{\Gamma _{k}}$\textit{\ of
the gauge theory with compact structure group }$G$\textit{.}
\end{theorem}

Notice that in the verification of the condition (\ref{5.6}) by (\ref{5.17})
what is important is the characters orthogonality relation. Therefore a
different set of $c\left( \lambda \right) $'s might be used. This would lead
to a different measure. The choice of which measure to choose would depend
on physical considerations in the small scale limit.

The $SU\left( 2\right) $ and $SU\left( 3\right) $ cases will now be analyzed
in detail

\subsubsection{$SU\left( 2\right) $}

Here%
\begin{eqnarray}
\Lambda ^{+} &=&\left\{ \lambda :2\lambda \in \mathbb{N},\lambda \geq
0\right\}  \notag \\
c\left( \lambda \right) &=&\frac{1}{2}\lambda \left( \lambda +1\right) 
\notag \\
d_{\lambda } &=&2\lambda +1  \label{5.19}
\end{eqnarray}%
\begin{equation}
K_{2}\left( g,\beta \right) =\sum_{\lambda \in \Lambda ^{+}}\left( 2\lambda
+1\right) e^{-\frac{1}{2}\lambda \left( \lambda +1\right) \beta }\frac{\sin
\left\{ \left( 2\lambda +1\right) \pi x\left( g\right) \right\} }{\sin
\left\{ \pi x\left( g\right) \right\} }  \label{5.20}
\end{equation}%
where $\pi x\left( g\right) $ is the angle coordinate of $g$ in a maximal
torus. It may be obtained from the $2\times 2$ matrix representation of $g$
by 
\begin{equation}
\pi x\left( g\right) =\cos ^{-1}\left( \frac{Tr\left( g\right) }{2}\right)
\label{5.21}
\end{equation}%
$K_{2}\left( g,\beta \right) $ may be rewritten as 
\begin{equation}
K_{2}\left( g,\beta \right) =2\left( 2\pi \right) ^{3/2}\frac{e^{\beta /8}}{%
\beta ^{3/2}}\sum_{n\in \mathbb{Z}}\frac{x\left( g\right) +2n}{\sin \left\{
\pi x\left( g\right) \right\} }\exp \left\{ -2\pi ^{2}\frac{\left( x\left(
g\right) +2n\right) ^{2}}{\beta }\right\}  \label{5.22}
\end{equation}%
One also sees that for small $\beta $ (small scales) the heat kernel density
is dominated by the $n=0$ term in the sum above and by field configurations
near $x\left( g\right) =0$. As in the $U\left( 1\right) $ case, at $\beta =0$
the \textquotedblleft densities\textquotedblright\ are no longer continuous
functions and the measure in the projective limit space is not absolutely
continuous with respect to the kinematical measure. There is however a
generalized density in $\mathcal{H}^{\ast }\left( \mathcal{A}\right) $.

The structure of the measure at small scales may now be compared with the
naive continuum limit of lattice theory, as discussed for example in \cite%
{Creutz}. There 
\begin{equation}
U=e^{icaA_{\mu }^{b}\tau _{b}}  \label{5.23}
\end{equation}
$c$ is a coupling constant, $a$ the lattice spacing, $A_{\mu }^{b}$ the
continuum gauge field and $\tau _{b}$ a Lie algebra basis. Then, for a
plaquette in the $1,2$ plane 
\begin{equation}
U_{\square }\left( A\right) =e^{icaA_{1}^{b}\left( x_{\mu }-\frac{1}{2}%
a\delta _{\mu 2}\right) \tau _{b}}e^{icaA_{2}^{b}\tau _{b}\left( x_{\mu }+%
\frac{1}{2}a\delta _{\mu 1}\right) }e^{-icaA_{1}^{b}\left( x_{\mu }+\frac{1}{%
2}a\delta _{\mu 2}\right) \tau _{b}}e^{-icaA_{2}^{b}\left( x_{\mu }-\frac{1}{%
2}a\delta _{\mu 1}\right) \tau _{b}}  \label{5.24}
\end{equation}
which for small $a$ leads to 
\begin{equation}
Tr\left( U_{\square }\left( A\right) \right) \simeq 2-\frac{c^{2}a^{4}}{4}%
F_{12}^{b}F_{12}^{b}  \label{5.25}
\end{equation}

On the other hand, from (\ref{5.21}), one knows that 
\begin{equation*}
TrU_{\square }=2\cos \left( \pi x\left( U_{\square }\right) \right)
\end{equation*}%
or $TrU_{\square }\simeq 2\left( 1-\frac{1}{2}\pi ^{2}x^{2}\left( U_{\square
}\right) \right) $ for small $x$. Comparing with (\ref{5.25}) one concludes
that 
\begin{equation*}
x^{2}\left( U_{\square }\right) =\frac{c^{2}a^{4}}{4\pi ^{2}}%
F_{12}^{b}F_{12}^{b}
\end{equation*}%
Therefore, for small $\beta $, the measure that uses (\ref{5.22}) as its
density coincides with the usual physical continuum measure.

In Fig.3 one plots the density (\ref{5.22}) for different values of $\theta
=\pi x$ and $\beta $. As in the $U\left( 1\right) $ case, one sees that for
small $\beta $ (small scales) the measure concentrates around $\theta =0$.

\begin{figure}[tbh]
\begin{center}
\psfig{figure=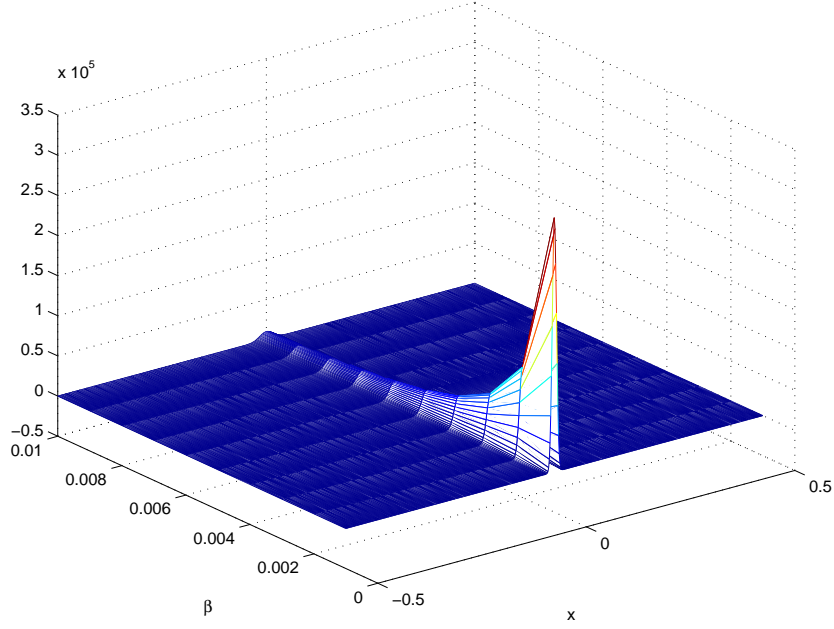,width=11truecm}
\end{center}
\caption{The $SU\left( 2\right) -$ heat
kernel density}
\end{figure}

\subsubsection{$SU\left( 3\right) $}

The irreducible representations of $SU\left( 3\right) $ are labelled by two
positive integers $\left( p,q\right) $ ($\left( 1,1\right) $ is the
one-dimensional representation, $\left( 2,1\right) =\underset{\backsim }{3}%
,\left( 1,2\right) =\underset{\backsim }{3}^{\ast }$, etc.). The eigenvalues
of the Laplacian and dimensions of the representations are 
\begin{eqnarray}
c\left( p,q\right) &=&\frac{1}{3}\left( p^{2}+q^{2}+pq\right) -1  \notag \\
d_{p,q} &=&\frac{1}{2}pq\left( p+q\right) \hspace{1cm}p,q=1,2,\cdots
\label{5.27}
\end{eqnarray}%
With $\left( \theta _{1}\left( g\right) ,\theta _{2}\left( g\right) \right) $
denoting the angle coordinates of $g$ in the maximal torus $diag\left\{ \exp
\left( i\theta _{1}\right) ,\exp \left( i\theta _{2}\right) ,\exp \left(
-i\left( \theta _{1}+\theta _{2}\right) \right) \right\} $, obtained from 
\begin{eqnarray}
\cos \theta _{1}\left( g\right) +\cos \theta _{2}\left( g\right) +\cos
\left( \theta _{1}-\theta _{2}\right) \left( g\right) &=&\textnormal{Re}Tr\left(
g\right)  \notag \\
\sin \theta _{1}\left( g\right) +\sin \theta _{2}\left( g\right) -\sin
\left( \theta _{1}-\theta _{2}\right) \left( g\right) &=&\textnormal{Im}Tr\left(
g\right)  \label{5.29}
\end{eqnarray}%
the $SU\left( 3\right) $ heat kernel is \cite{Baaquie} 
\begin{eqnarray}
&&K_{3}\left( g,\beta \right)  \notag \\
&=&\frac{1}{s\left( \theta _{1},\theta _{2}\right) }\sum_{l=-\infty
}^{\infty }\sum_{m=-\infty }^{\infty }\left\{ \theta _{1}-\theta _{2}+2\pi
\left( l-m\right) \right\} \left\{ \theta _{1}+2\theta _{2}+2\pi \left(
l+2m\right) \right\} \times  \notag \\
&&\left\{ 2\theta _{1}+\theta _{2}+2\pi \left( 2l+m\right) \right\} \exp
\left\{ -\frac{1}{2\beta }\left[ \left( \theta _{1}+2\pi l\right)
^{2}+\left( \theta _{2}+2\pi m\right) ^{2}+\left( \theta _{1}+2\pi l\right)
\left( \theta _{2}+2\pi m\right) \right] \right\}  \notag \\
&&  \label{5.30}
\end{eqnarray}%
with 
\begin{equation}
s\left( \theta _{1},\theta _{2}\right) =8\sin \frac{1}{2}\left( \theta
_{1}-\theta _{2}\right) \sin \frac{1}{2}\left( 2\theta _{1}+\theta
_{2}\right) \sin \frac{1}{2}\left( \theta _{1}+2\theta _{2}\right)
\label{5.31}
\end{equation}%
That the choice $p\left( g\right) =K_{3}\left( g,\beta \right) $ and $%
p^{\prime }\left( g\right) =K_{3}\left( g,\frac{\beta }{4}\right) $, for
each plaquette subdivision, satisfies the consistency condition (\ref{5.6})
follows from the general result (Theorem 3).

Now one checks the consistency of the result with the physical small scale
(small $\beta $) limit. Inspection of (\ref{5.30}) leads to the conclusion
that for $\beta =0$ the sums become dominated by the $l=m=0$ term and the
neighborhood $\left( \theta _{1}\thickapprox 0,\theta _{2}\thickapprox
0\right) $. Then 
\begin{equation*}
K_{3}\left( g,\beta \right) \underset{\beta \rightarrow 0}{\longrightarrow }%
\exp \left\{ -\frac{1}{2\beta }\left( \theta _{1}^{2}+\theta _{2}^{2}+\theta
_{1}\theta _{2}\right) \right\}
\end{equation*}%
From (\ref{5.29}) one also sees that for small $\theta _{1},\theta _{2}$%
\begin{equation*}
3-\theta _{1}^{2}-\theta _{2}^{2}-\theta _{1}\theta _{2}\backsimeq Tr\left(
g\right)
\end{equation*}%
which, equating with (\ref{5.25}), leads to 
\begin{equation*}
\theta _{1}^{2}+\theta _{2}^{2}+\theta _{1}\theta _{2}\backsimeq 1-\frac{%
c^{2}a^{4}}{4}F_{12}^{b}F_{12}^{b}
\end{equation*}%
That is, up to an irrelevant constant factor (to be absorbed in the measure
normalization), one also obtains in the $SU(3)$ case the usual physical
measure in the $\beta \rightarrow 0$ limit. Therefore, both for $SU\left(
2\right) $ and $SU\left( 3\right) $, one may take the measures so
constructed as a definition of the Yang-Mills measure.

For small $\beta $ the measure is always concentrated in the neighborhood $%
\left( \theta _{1}\thickapprox 0,\theta _{2}\thickapprox 0\right) $.

\section{Gauge measure and the strata}

Let $S_{A}$ be the stabilizer (isotropy group) of a generalized connection $%
A\in \mathcal{A}$%
\begin{equation}
S_{A}=\left\{ \gamma \in \mathcal{G}:\gamma A=A\right\}  \label{7.1}
\end{equation}%
The action of the gauge group $\mathcal{G}$ on $\mathcal{A}$ leads to a
stratification of $\mathcal{A}$ corresponding to the classes of equivalent 
\textit{orbits} $\left\{ gA;g\in \mathcal{G}\right\} $. The \textit{stratum} 
$\Sigma \left( A\right) $ of $A$ is the set of connections having isotropy
groups $\mathcal{G}-$conjugated to that of $A$%
\begin{equation}
\Sigma \left( A\right) =\left\{ B\in \mathcal{A}:\exists \gamma \in \mathcal{%
G}:S_{B}=\gamma S_{A}\gamma ^{-1}\right\}  \label{7.2}
\end{equation}%
The \textit{configuration space of the gauge theory} is the quotient space $%
\mathcal{A}/\mathcal{G}$ and therefore a stratum is the set of points in $%
\mathcal{A}/\mathcal{G}$ that correspond to orbits with conjugated isotropy
groups.

When $G$ is a compact group the stratification is topologically regular. The
map that, to each orbit, assigns the conjugacy class of its isotropy group
is called the \textit{type}. The set of strata carries a partial ordering of
types, $\Sigma _{\tau }\subseteq \Sigma _{\tau ^{^{\prime }}}$ with $\tau
\leq \tau ^{\prime }$ if there are representatives $S_{\tau }$ and $S_{\tau
^{\prime }}$ of the isotropy groups such that $S_{\tau }\supseteq S_{\tau
^{\prime }}$. The maximal element in the ordering of types is the class of
the center $Z(G)$ of $G$ and the minimal one is the class of $G$ itself.
Furthermore $\cup _{t\geq \tau }\Sigma _{t}$ is open and $\Sigma _{\tau }$
is open in the relative topology in $\cup _{t\leq \tau }\Sigma _{t}$.

Because the isotropy group of a connection is isomorphic to the centralizer
of its holonomy group, the strata are in one-to-one correspondence with the
Howe subgroups of $G$, that is, the subgroups that are centralizers of some
subset in $G$. Given an holonomy group $H_{\tau }$ associated to a
connection $A$ of type $\tau $, the stratum of $A$ is classified by the
conjugacy class of the isotropy group $S_{\tau }$, that is, the centralizer
of $H_{\tau }$ 
\begin{equation}
S_{\tau }=Z\left( H_{\tau }\right)  \label{7.3}
\end{equation}%
An important role is also played by the centralizer of the centralizer 
\begin{equation}
H_{\tau }^{\prime }=Z\left( Z\left( H_{\tau }\right) \right)  \label{7.4}
\end{equation}%
that contains $H_{\tau }$ itself. If $H_{\tau }^{\prime }$ is a proper
subgroup of $G$, the connection $A$ reduces locally to the subbundle $%
P_{\tau }=\left( M,H_{\tau }^{\prime }\right) $. Global reduction depends on
the topology of $M$, but it is always possible if $P$ is a trivial bundle. $%
H_{\tau }^{\prime }$ is the structure group of the \textit{maximal subbundle}
associated to type $\tau $. Therefore the types of strata are also in
correspondence with types of reductions of the connections to subbundles. If 
$S_{\tau }$ is the center of $G$ the connection is called \textit{irreducible%
}, all others are called \textit{reducible}. The stratum of the irreducible
connections is called the \textit{generic stratum}.

Now, for $G=SU\left( 2\right) $ and $SU\left( 3\right) $ one describes the
strata and how they stand in relation to the measures defined before. In $%
G=SU(2)$, the isotropy groups (equivalently, the centralizers of the
holonomy) and the structure groups of the maximal subbundles are : 
\begin{equation}
\begin{tabular}{|c|c|c|}
\hline
& $S_{A}$ & $H_{A}^{^{\prime }}$ \\ \hline
1 & $\mathbb{Z}_{2}$ & $SU(2)$ \\ \hline
2 & $U\left( 1\right) $ & $U\left( 1\right) $ \\ \hline
3 & $SU\left( 2\right) $ & $\mathbb{Z}_{2}$ \\ \hline
\end{tabular}
\label{7.5}
\end{equation}%
There are three strata. Stratum 1 is the generic stratum. The other two are
reducible strata.

A $SU\left( 2\right) $ transformation may be parametrized by%
\begin{equation}
\exp \left( -i\frac{\theta }{2}\overset{\rightarrow }{n}\cdot \overset{%
\rightarrow }{\sigma }\right) =\boldsymbol{1}\cos \frac{\theta }{2}+i\left( 
\overset{\rightarrow }{n}\cdot \overset{\rightarrow }{\sigma }\right) \sin 
\frac{\theta }{2}  \label{7.6}
\end{equation}%
Geometrically, the $SU\left( 2\right) $ group may be pictured as a sphere of
radius $4\pi $, with all the points at radial distance $2\pi $ identified to 
$\left( 
\begin{array}{cc}
-1 & 0 \\ 
0 & -1%
\end{array}%
\right) $ and the points at radius $4\pi $ identified with the center of the
sphere. The reducible stratum 3 corresponds to a $\mathbb{Z}_{2}-$bundle,
that is, to homomorphisms of the loops to a two point space $\left( \theta
=0,\theta =2\pi \right) $. Each reducible stratum of type 2 is a $U\left(
1\right) -$bundle corresponding to homorphisms of the loops to the $SU\left(
2\right) $ transformations along one radius (fixed $\overset{\rightarrow }{n}
$, variable $\theta $). Because adjoint transformations transform any radius
into any other, all $U\left( 1\right) -$bundles are equivalent and represent
the same gauge configurations. Finally, the (generic) stratum 1 corresponds
to homomorphisms to arbitrary $SU\left( 2\right) $ transformations.

From (\ref{5.21}) one sees that the intensity of the measure (\ref{5.20})
only depends on the first term $\boldsymbol{1}\cos \frac{\theta }{2}$ in the
parameterization (\ref{7.6}). Therefore all strata approach the small $%
x\left( g\right) $ region where the measure is peaked and therefore they all
are expected to be relevant in the physical behavior of the gauge theory.

For $G=SU(3)$ the isotropy groups and the structure groups of the maximal
subbundles are : 
\begin{equation}
\begin{tabular}{|c|c|c|}
\hline
& $S_{A}$ & $H_{A}^{^{\prime }}$ \\ \hline
1 & $\mathbb{Z}_{3}$ & $SU\left( 3\right) $ \\ \hline
2 & $U\left( 1\right) $ & $U\left( 2\right) $ \\ \hline
3 & $U(1)\times U(1)$ & $U(1)\times U(1)$ \\ \hline
4 & $U\left( 2\right) $ & $U\left( 1\right) $ \\ \hline
5 & $SU\left( 3\right) $ & $\mathbb{Z}_{3}$ \\ \hline
\end{tabular}
\label{7.7}
\end{equation}%
There are five strata. Stratum 1 is the generic stratum. All others are
reducible strata. To find out their relevance in the framework of the
measure (\ref{5.30}) one uses the following parametrization \cite{Macfarlane}
for an arbitrary $SU\left( 3\right) $ transformation,%
\begin{equation}
U=u_{0}+iu_{k}\lambda _{k}  \label{7.8}
\end{equation}%
where $u_{k}=\alpha a_{k}+\beta d_{ijk}a_{i}a_{j}$ for an octet vector $%
a_{k} $, with $u_{0},\alpha $ and $\beta $ being functions of the invariants%
\begin{equation}
\begin{array}{lll}
I_{2}\left( a\right) & = & a_{i}a_{i} \\ 
I_{3}\left( a\right) & = & d_{ijk}a_{i}a_{j}a_{k}%
\end{array}
\label{7.9}
\end{equation}
which can be built from the vector $a_{k}$. Then%
\begin{equation}
\begin{array}{lll}
u_{0} & = & \frac{1}{3}\sum_{n}e^{i\varphi _{n}} \\ 
\alpha & = & -\sum_{n}\varphi _{n}e^{i\varphi _{n}}\left( 3\varphi
_{n}^{2}-I_{2}\right) ^{-1} \\ 
\beta & = & -\sum_{n}e^{i\varphi _{n}}\left( 3\varphi _{n}^{2}-I_{2}\right)
^{-1}%
\end{array}
\label{7.10}
\end{equation}%
with%
\begin{equation}
\varphi _{n}=2\left( \frac{I_{2}}{3}\right) ^{\frac{1}{2}}\cos \frac{1}{3}%
\left( 2\pi n+\cos ^{-1}\left( \sqrt{3}I_{3}I_{2}^{-\frac{3}{2}}\right)
\right) ;\hspace{0.45cm}n=1,2,3  \label{7.11}
\end{equation}

As before, one sees from (\ref{5.29}) that the intensity of the measure only
depends on $\textnormal{Re}u_{0}$ and $\textnormal{Im}u_{0}$, whereas the choice of the
subbundle for each stratum depends on the choice of the $u_{k}$
coefficients. Therefore there is for all strata a range of parameters that
approaches the region where the measure is peaked.

\section{Remarks and conclusions}

1) The Cartesian product point of view, in the construction of the
projective limit $\mathcal{A}=\underset{\longleftarrow }{\lim }\mathcal{A}%
_{\Gamma }$ and of the triplet $\mathcal{H}\left( \mathcal{A}\right) \subset
L^{2}\left( \mathcal{A},d\nu \right) \mathcal{\subset H}^{\ast }\left( 
\mathcal{A}\right) $,\ means that a consistent framework is obtained for the
description of gauge theories at all length scales. In this setting the
lattice structure, underlying the Cartesian product, is not an approximation
scheme but a framework to characterize the theory at the several length
scales.

2) For each $\mathcal{A}_{\Gamma }=Hom\left( E\left( \Gamma \right)
,G\right) \sim G^{\#E\left( \Gamma \right) }$, the interaction measure $%
\frac{1}{Z}p\left( \mathcal{A}_{\Gamma }\right) \left( d\mu _{H}\right)
^{\#E\left( \Gamma \right) }$, constructed in Section 4 is absolutely
continuous with respect the kinematical measure. However when $\beta
\rightarrow 0$ the "density"\ ceases to be a continuous function. Therefore,
for the full projective limit, the interaction measure is not absolutely
continuous with respect to the kinematical measure. That such a result was
to be expected, follows also from the analysis of Fleischhak \cite%
{Fleischhack1} \cite{Fleischhack3} who, starting from very general
conditions concluded that in gauge theories there is a "breakdown of the
action method". However, in the gauge triplet framework, one may consider
that a generalized density exists in $\mathcal{H}^{\ast }\left( \mathcal{A}%
\right) $.

3) The fact that the small $\beta $ component of the interaction measures
coincides with the usual small distance representations of the abelian and
the Yang-Mills measure, mean that they will inherit the same qualitative
physical properties. In particular, the absence of a mass gap in the abelian
case and, for the non-abelian case, the same qualitative properties as
obtained, for example, from asymptotic dynamics \cite{Vilela3}, namely the
fact that either there is spontaneous violation of the symmetry or all
asymptotic states are color singlets.

4) As to the role of non-generic strata in gauge theories, the fact that
there is for all strata a range of parameters that approaches the region
where the measure is peaked, emphasizes the importance of these strata for
the structure of low-lying excitations. A similar result \cite{Vilela2} had
been obtained using a ground-state approximation \cite{Vilela} \cite{Vilela4}
for the non-abelian ground-state.


\begin{thebibliography}{99}
\bibitem{Ashtekar1} A. Ashtekar and C. J. Isham; \textit{Representations of
the holonomy algebras of gravity and nonabelian gauge theories}, Class.
Quant. Grav. 9 (1992) 1433-1468.

\bibitem{Ashtekar2} A. Ashtekar and J. Lewandowski; \textit{Differential
geometry on the space of connections via graphs and projective limits}, J.
Geom. Phys. 17 (1995) 191-230.

\bibitem{Ashtekar3} A. Ashtekar and J. Lewandowski; \textit{Projective
techniques and functional integration for gauge theories}, J.Math. Phys. 36
(1995) 2170-2191.

\bibitem{Marolf} D. Marolf and J. M. Mour\~{a}o; \textit{On the support of
the Ashtekar-Lewandowski measure}, Commun. Math. Phys. 170 (1995) 583-606.

\bibitem{Ashtekar0} A. Ashtekar and J. Lewandowski; \textit{Representation
theory of analytic holonomy C}$^{*}$\textit{\ algebras},
(arXiv:gr-qc/9311010) in \textit{Knots and Quantum Gravity} (J. Baez, Ed.)
Oxford Univ. Press, Oxford 1994.

\bibitem{Baez1} J. Baez and S. Sawin; \textit{Functional integration on
spaces of connections}, J. Funct. Anal. 150 (1997) 1-26.

\bibitem{Fleichhack1} C. Fleischhack; \textit{Stratification of the
generalized gauge orbit space}, Comm. Math. Phys. 214 (2000) 607-649.

\bibitem{Fleischhack2} C. Fleischhack; \textit{On the Structure of Physical
Measures in Gauge Theories}, arXiv:math-ph/0107022.

\bibitem{Fleichhack2} C. Fleischhack; \textit{Hyphs and the
Ashtekar-Lewandowski measure}, J. Geom. Phys. 45 (2003) 231-251.

\bibitem{Kisynski} J. Kisynski;\textit{\ On the generation of tight measures}%
, Studia Math. 30 (1968) 141-151.

\bibitem{Maurin} K. Maurin; \textit{General eigenfunction expansions and
unitary representations of topological groups}, PWN - Polish Scient. Publ.,
Warszawa 1968.

\bibitem{Deck} T. Deck; \textit{Hida distributions on compact Lie groups},
Inf. Dim. Anal., Quantum Prob. and Rel. Topics 3 (2000) 337-362.

\bibitem{Gelfand} I. M. Gelfand and G. E. Shilov; \textit{Generalized
functions}, vol.2, Academic Press, NY 1968.

\bibitem{Creutz} M. Creutz; \textit{Quarks, gluons and lattices}, Cambridge
U. P., Cambridge 1983.

\bibitem{Baaquie} B. E. Baaquie; \textit{Character functions of SU(3)}, J.
Phys. A:\ Math. Gen. 21 (1988) 2651-2656.

\bibitem{Macfarlane} A. J. Macfarlane, A. Sudbery and P. H. Weisz; \textit{%
On Gell-Mann's }$\lambda $\textit{-matrices, d- and f-tensors, octets and
parametrization of SU(3)}, Commun. Math. Phys. 11 (1968) 77-90.

\bibitem{Fleischhack1} C. Fleischhack;\textit{\ On the structure of physical
measures in gauge theories, }arXiv:math-ph/0107022.

\bibitem{Fleischhack3} C. Fleischhack; \textit{On the support of physical
measures in gauge theories}, arXiv:math-ph/0109030.

\bibitem{Vilela3} R. Vilela Mendes; \textit{Asymptotic dynamics for gauge
theories}, Phys. Rev. D18 (1978) 4726-4736.

\bibitem{Vilela2} R. Vilela Mendes; \textit{Stratification of the orbit
space in gauge theories. The role of nongeneric strata}, J. Phys. A: Math.
Gen. 37 (2004) 11485-11498.

\bibitem{Vilela} R. Vilela Mendes; \textit{Stochastic processes and the
non-perturbative structure of the QCD vacuum}, Z. Phys. C - Particles and
Fields 54 (1992)\ 273-281.

\bibitem{Vilela4} R. Vilela Mendes; \textit{Path-integral estimates of
ground-state functionals}, Proceedings of the International Conference on
Mathematical Analysis of Random Phenomena, pags. 169-177, World Scientific
2006.
\end{thebibliography}
\end{document}